\documentclass[pra,twocolumn,superscriptaddress]{revtex4-2}
\usepackage[T1]{fontenc}
\usepackage{amsmath}
\usepackage{amssymb}
\usepackage{hyperref}

\newcommand{\ketbra}[1]{| #1\rangle \langle #1|}
\newcommand{\va}[1]{\ensuremath{(\Delta#1)^2}}
\newcommand{\ex}[1]{\ensuremath{\langle{#1}\rangle}}
\newcommand{\qed}{\ensuremath{\hfill \blacksquare}}
\newcommand{\trace}{{\rm Tr}}

\newcommand{\EQ}[1]{Eq.~\eqref{#1}}
\newcommand{\EQS}[1]{Eqs.~\eqref{#1}}

\newcommand{\SEC}[1]{Sec.~\ref{#1}}

\newcommand{\REF}[1]{Ref.~\cite{#1}}
\newcommand{\REFS}[1]{Refs.~\cite{#1}}

\newcommand{\THM}[1]{Theorem~\ref{#1}}

\newcommand{\DEFTHM}[1]{{\bf Theorem \refstepcounter{theorem}\thetheorem\label{#1}.} }
\newcounter{theorem}
\newcommand{\LEMMA}[1]{Lemma~\ref{#1}}

\newcommand{\DEFLEMMA}[1]{{\bf Lemma \refstepcounter{lemma}\thelemma\label{#1}.} }
\newcounter{lemma}
\newcommand{\DEFINITION}[1]{Definition~\ref{#1}}
\newcommand{\DEFINITIONS}[1]{Definitions~\ref{#1}}
\newcommand{\DEFDEFINITION}[1]{{\bf Definition \refstepcounter{definition}\thedefinition\label{#1}.} }
\newcounter{definition}

\begin{document}

\title{Relations between different definitions of the quantum Wasserstein distance}

\author{G\'eza T\'oth}
\email{toth@alumni.nd.edu}
\affiliation{Theoretical Physics, University of the Basque Country UPV/EHU,  
48080 Bilbao, Spain}
\affiliation{EHU Quantum Center, University of the Basque Country UPV/EHU, 
48940 Leioa, 
Spain}
\affiliation{Donostia International Physics Center DIPC,  
20018 San Sebasti\'an, Spain}
\affiliation{IKERBASQUE, Basque Foundation for Science, 48009 Bilbao, Spain}
\affiliation{HUN-REN Wigner Research Centre for Physics,  
1525 Budapest, Hungary}

\author{J\'ozsef Pitrik}
\email{pitrik@math.bme.hu}
\affiliation{Department of Analysis and Operations Research, Institute of Mathematics, Budapest University of Technology and Economics, 
1111 Budapest, Hungary}
\affiliation{HUN-REN 
R\'enyi Institute of Mathematics, 
1053 Budapest, Hungary}
\affiliation{HUN-REN Wigner Research Centre for Physics,  
1525 Budapest, Hungary}

\begin{abstract}
The quantum Wasserstein distances defined by Golse, Mouhot, Paul, and Caglioti and by De Palma and Trevisan coincide for qubits when a single operator appears in the cost function. As a consequence, the square of the self-distance equals the Wigner-Yanase skew information in this case. Moreover, for finite-dimensional systems of dimension $d\ge2$ and any number of operators, we show that the Golse-Mouhot-Paul-Caglioti self-distance is bounded from below by the De Palma-Trevisan self-distance.
\end{abstract}

\date{\today}

\maketitle

\section{Introduction}

Quantum optimal transport has led to several inequivalent extensions of classical optimal-transport concepts to the quantum setting. Early semiclassical constructions were introduced by \.Zyczkowski and S\l omczy\'nski in connection with quantum chaos \cite{Zyczkowski1998TheMonge,Zyczkowski2001TheMonge,Bengtsson2006Geometry}, while Biane and Voiculescu developed an approach related to free probability \cite{Biane2011Free}. Dynamical formulations were subsequently studied by Carlen, Maas, Datta, and Rouz\'e \cite{CarlenMaas2014Analog,CarlenMaas2017Gradient,CarlenMaas2020Non-commutative,DattaRouze2019Concentration,DattaRouze2020Relating}. A different, static construction was introduced by Caglioti, Golse, Mouhot, and Paul \cite{Golse2016On,Golse2017The,Golse2018Wave,Golse2018TheQuantum,Caglioti2020Quantum,Caglioti2023Towards}. Lafleche related this construction to negative Sobolev norms \cite{lafleche2023quantumoptimaltransportweak}. Further approaches include the quantum-channel formulation of De Palma and Trevisan \cite{DePalma2021Quantum,DePalma2024QuantumOptimal}, and the quantum earth mover's distance of De Palma, Marvian, Trevisan, and Lloyd, i.e., the quantum Wasserstein distance of order 1 \cite{DePalma2021TheQuantum}, and the Wasserstein-type distance based on an antisymmetric cost function and the SWAP-fidelity introduced by Bistro\'n, Cole, Eckstein, Friedland, and \.Zyczkowski \cite{Friedland2022Quantum,Cole2023OnQuantum,Bistron2023Monotonicity}. 

These developments have produced several notions of quantum Wasserstein distance with properties that differ markedly from their classical counterparts \cite{Zyczkowski1998TheMonge,Zyczkowski2001TheMonge,Bengtsson2006Geometry,Golse2016On,Golse2017The,Golse2018Wave,Golse2018TheQuantum,DePalma2021Quantum,DePalma2021TheQuantum,Caglioti2020Quantum,Caglioti2023Towards,Geher2023Quantum,Li2025Wasserstein}. For reviews see \REFS{Trevisan2025Quantum,Beatty2026Wasserstein}. They have also found applications, for instance, in quantum machine learning \cite{Kiani2022Learning,malyshev2026distancelearningprojectivemeasurements}. Of particular relevance here is the quantum-channel construction associated with a  given set of operators $\{H_n\}_{n=1}^N$ \cite{DePalma2021Quantum,DePalma2024QuantumOptimal}. Its squared self-distance for a state $\varrho$ is equal to the sum of the corresponding Wigner--Yanase skew informations $I{\varrho}^{\rm WY}(H_n)$ of the quantum state \cite{DePalma2021Quantum,Wigner1963INFORMATION}. This formulation has been applied to the comparison of ground-state phases in spin systems \cite{Camacho2025Critical}, generalized to $p$-Wasserstein distances \cite{bunth2025wassersteindistancesdivergencesorder}, and studied from the point of view of duality   \cite{bunth2025strongkantorovichdualityquantum}. A related modified distance has vanishing self-distance  \cite{DePalma2021Quantum} and satisfies the triangle inequality \cite{Bunth2024MetricProperty,bunth2025wassersteindistancesdivergencesorder,wirth2025triangleinequalityquantumwasserstein}.

Another line of work considers optimizations over bipartite separable states. For a single operator, the squared self-distance equals one quarter of the quantum Fisher information \cite{Toth2023QuantumWasserstein}, connecting this construction to entanglement theory \cite{Horodecki2009Quantum,Guhne2009Entanglement,Friis2019}  and  quantum metrology \cite{Giovannetti2004Quantum-Enhanced,Paris2009QUANTUM,Demkowicz-Dobrzanski2014Quantum,Pezze2014Quantum,Toth2014Quantum,Pezze2018Quantum}. The same framework extends the pure-state distance based on the channel formalism to mixed states through an optimization over separable states with fixed marginals \cite{Toth2023QuantumWasserstein}. Beatty and Stilck-Fran\c{c}a considered the same separable-state optimization, with the distance based on the channel formalism replaced by the trace distance or by other pure-state distances, in which case the resulting self-distance can vanish \cite{Beatty2026Order}. 
Borsoni has generalized the approach to convex sets different from separable states \cite{borsoni2026foldedoptimaltransportapplication}.

Various definitions based on optimizations over separable states, including the distance coming from the SWAP-fidelity, have been related to each other and to the Uhlmann-Jozsa fidelity. Modified De Palma–Trevisan-type separable distances have been introduced with vanishing self-distance. For qubits, the convex roof of the squared trace-distance between pure states equals one minus the fidelity, and therefore its square root defines a metric. Triangle inequalities also hold when an intermediate state is pure, and for a larger family of costs generated by observables $H_n,$ in every dimension $d\ge 2$ \cite{Toth2025QuantumWasserstein}. Miller has recently proved the triangle inequality in full generality for the separable distance arising from the squared trace-distance cost \cite{miller2026metricityseparablequantumoptimal}. By the equivalences described above, this also establishes metricity for the separable SWAP-fidelity distance and, for a complete set of pairwise orthogonal observables $H_n,$ for the corresponding modified De Palma–Trevisan-type separable distances.

Since there are several possible definitions of the quantum Wasserstein distance, the question arises: Is it possible to connect these approaches to each other? As we have mentioned, there are several possible definitions of the quantum Wasserstein distance of order 2. Two of them are the Golse-Mouhot-Paul-Caglioti distance \cite{Golse2016On,Caglioti2023Towards,Golse2018TheQuantum,Golse2017The,Golse2018Wave,Caglioti2020Quantum} and the De Palma-Trevisan distance \cite{DePalma2021Quantum}, both based on optimizations over bipartite states with prescribed marginals. In this paper, we look for relations between these two definitions. It has been shown that the separable quantum Wasserstein distance presented in \REF{Toth2023QuantumWasserstein} is an upper bound on both of these distances. We will identify cases in which the two quantities are equal. We will also derive inequalities between their squared self-distances.

Our paper is organized as follows. In \SEC{sec:background}, we summarize the definitions used in our paper. In \SEC{sec:main}, we present our main results. In \SEC{sec:cons}, we discuss some consequences of our findings for computing the energy minimum for spin systems.

\section{Background}
\label{sec:background}

Before presenting our results, let us summarize the definitions of the quantum Wasserstein distance relevant for our article.

\DEFDEFINITION{def:DGMCP}Golse, Mouhot, Paul, and Caglioti  defined the square of the distance between two quantum states described by the density matrices $\varrho$ and $\sigma$ as \cite{Golse2016On,Caglioti2023Towards,Golse2018TheQuantum,Golse2017The,Golse2018Wave,Caglioti2020Quantum}
\begin{align}
\begin{split}
D_{\rm GMPC}(\varrho,\sigma)^2\quad\quad\quad\\=\frac 1 2
\min_{ \varrho_{12}} \sum_{n=1}^N\;&
\trace[(H_n\otimes \openone-\openone\otimes H_n)^2  \varrho_{12} ],\\
\textrm{s.t. }&
\varrho_{12}\in\mathcal D,\label{eq:GMPC_distance}\\
& {\rm Tr}_2(\varrho_{12})=\varrho,\\
& {\rm Tr}_1(\varrho_{12})=\sigma,
\end{split}
\end{align}
where $H_1, H_2, ...,H_N$ are Hermitian operators \footnote{We use the convention that in the expression within the trace in \EQ{eq:GMPC_distance}, $H_n\otimes \openone$ denotes an operator in which $H_n$ acts on subsystem $1$ and $\openone$ acts on subsystem $2$}, while $\mathcal D$ is the set of density matrices, i.e., Hermitian matrices fulfilling
\begin{equation}
\varrho_{12} =\varrho_{12}^\dagger, {\rm Tr}(\varrho_{12})=1, \varrho_{12}\ge 0.
\end{equation}

The Wasserstein distance has also been defined in a different way. 

\DEFDEFINITION{def:D}The square of the distance  is given by De Palma and Trevisan as \cite{DePalma2021Quantum}
\begin{align}
\begin{split}
D_{\rm DPT}(\varrho,\sigma)^2=\frac 1 2
\min_{ \varrho_{12} }\sum_{n=1}^N\;&
\trace[(H_n^T\otimes \openone-\openone\otimes H_n)^2  \varrho_{12} ],\\
\textrm{s.t. }&
\varrho_{12}\in\mathcal D,\\
& {\rm Tr}_2(\varrho_{12})=\varrho^T,\\
& {\rm Tr}_1(\varrho_{12})=\sigma,\label{eq:distance}
\end{split}
\end{align}
where $A^T$ denotes the matrix transpose of $A.$ In this approach, there is a bipartite density matrix $\varrho_{12},$ called coupling, corresponding to any transport map between $\varrho$ and $\sigma,$ and vice versa, there is a transport map corresponding to any coupling \cite{DePalma2021Quantum}. Moreover, it has been shown that the square of the self-distance of a state  is given by \cite{DePalma2021Quantum}
\begin{equation}
D_{\rm DPT}(\varrho,\varrho)^2=\sum_{n=1}^N I^{\rm WY}_\varrho(H_n),\label{eq:DrhoI}
\end{equation}
where the Wigner-Yanase skew information is defined as \cite{Wigner1963INFORMATION}
\begin{equation}
I^{\rm WY}_{\varrho}(H)={\rm Tr}(H^2\varrho)-{\rm Tr}(H\sqrt{\varrho}H\sqrt{\varrho}).\label{eq:WY}
\end{equation}
This profound result connects the square of the self-distance to a fundamental quantity in quantum physics and estimation theory \cite{Giovannetti2004Quantum-Enhanced,Paris2009QUANTUM,Demkowicz-Dobrzanski2014Quantum,Pezze2014Quantum,Toth2014Quantum,Pezze2018Quantum}. It is also important that the squared self-distance is obtained as an explicit formula. It has also been shown that the inequality
\begin{equation}
D_{\rm DPT}(\varrho,\sigma)^2-\frac 1 2 \left[D_{\rm DPT}(\varrho,\varrho)^2+D_{\rm DPT}(\sigma,\sigma)^2\right]\ge0.\label{eq:dptdiv}
\end{equation}
holds \cite{DePalma2021Quantum}.

The question arises: What is the relation between these two quantities, $D_{\rm GMPC}(\varrho,\sigma)$ and $D_{\rm DPT}(\varrho,\sigma)$? Can we obtain information on the self-distance of the GMPC distance, using the results on the DPT distance? So far, based on results about the separable Wasserstein distance, it has been known that for the GMPC distance for $N=1$  \cite{Toth2023QuantumWasserstein}
\begin{equation}
D_{\rm GMPC}(\varrho,\varrho)^2\le \frac1 4 F_Q[\varrho,H_1]
\end{equation}
holds  \cite{Toth2023QuantumWasserstein}, where the quantum Fisher information is defined as 
\begin{equation}
\label{eq:FQ}
F_Q[\varrho,{H}]=2\sum_{k,l}\frac{(\lambda_{k}-\lambda_{l})^{2}}{\lambda_{k}+\lambda_{l}}\vert \langle k \vert {H} \vert l \rangle \vert^{2},
\end{equation}
and the density matrix has the eigendecomposition
\begin{equation}\label{eq:rho_eigdecomp}
\varrho=\sum_{k}\lambda_k \ketbra{k}.
\end{equation}
For $N\ge2,$ the following bound is obtained \cite{Toth2023QuantumWasserstein}
\begin{align}
\begin{split}
D_{\rm GMPC}(\varrho,\varrho)^2& \le \inf_{\{p_k,\Psi_k\}}\sum_k p_k \sum_{n=1}^N (\Delta H_n)^2_{\Psi_k},
\end{split}
\end{align}
where the optimization is over decompositions fulfilling 
\begin{equation}
\varrho=\sum_k p_k \ketbra{\Psi_k}.
\end{equation}
We can also bound the Wasserstein distance with the cost computed for product states \cite{Toth2023QuantumWasserstein}
\begin{align}
\begin{split}
&D_{\rm GMPC}(\varrho,\sigma)^2\\
&\quad\le\frac1 2 \sum_{n=1}^N\left[\va{H_n}_{\varrho}+\va{H_n}_{\sigma}+(\ex{H_n}_{\varrho}-\ex{H_n}_{\sigma})^2\right].
\end{split}
\end{align}

We can also find relations between the two distances when the Hamiltonians have real matrix elements, with a stronger statement when $\varrho$ is real as well. It is clear by looking at the Definitions~\ref{def:DGMCP} and \ref{def:D} that if all $H_n$ are real (i.e., $H_n=H_n^T$ for all $n$) then
\begin{equation}
D_{\rm GMPC}(\varrho,\sigma)^2=D_{\rm DPT}(\varrho^T,\sigma)^2\label{eq:GMPCDPT}
\end{equation}
holds. If all $H_n$ are real, and additionally $\varrho$ is also real and hence $\varrho=\varrho^T,$ we have
\begin{equation}
D_{\rm GMPC}(\varrho,\sigma)^2=D_{\rm DPT}(\varrho,\sigma)^2.\label{eq:GMPCDPT2}
\end{equation}
After these considerations, we will examine the relation between the two distances for qubits, and subsequently for higher-dimensional systems.

\section{Main results}
\label{sec:main}

In this section, we will present our main results. We will prove that  the two distances are equal to each other for arbitrary qubit states $\varrho$ and $\sigma$ when a single operator $H_1$ is considered. We will also relate the two self-distances to each other for quantum systems of dimension $d\ge2.$

In order to proceed, we will need the following lemma.

\DEFLEMMA{lemma:pt}For a single qubit Hermitian operator $H$ and a single-qubit state $\varrho,$ there is always a unitary $U$ such that
\begin{align}
\begin{split}
H'&=U H U^\dagger,\\
\varrho'&=U\varrho U^\dagger,
\end{split}
\end{align}
are both real matrices.

{\it Proof.} Let us consider the eigendecomposition of $\varrho$
\begin{equation}
\varrho=V D V^\dagger,
\end{equation}
where $D$ is a diagonal matrix with real elements and  $V$ is unitary. Then, let us apply the unitary $V^\dagger$ to both $\varrho$ and $H.$ The density matrix
\begin{equation}
\varrho_2= V^\dagger \varrho V=D
\end{equation}
is a diagonal matrix with real elements. For this matrix
\begin{equation}
\ex{\sigma_x}=\ex{\sigma_y}=0 \label{eq:xz}
\end{equation}
holds. Then, we define the rotated Hermitian operator as 
\begin{equation}
H_2=  V^\dagger H V.
\end{equation}

We can write $H_2$ as
\begin{equation}
H_2=c_2\openone+\vec{h}_2 \cdot \vec{\sigma},
\end{equation}
where $\sigma=(\sigma_x,\sigma_y,\sigma_z)^T$ is a vector of the Pauli spin matrices, $\vec{h}_2$ is a vector of real numbers, and $c_2$ is a real number. 

The density matrix can be written similarly as 
\begin{equation}
\varrho_2=\frac{1}{2}\openone+\frac1 2 \vec{r}_2\cdot\vec{\sigma},
\end{equation}
where $\vec{r}_2$ is a vector of real numbers with $|\vec{r}_2|\le 1.$ Based on \EQ{eq:xz}, $\vec{r}_2$ is parallel to the $z$-axis, or $\vec{r}_2$ is the null vector, if $\varrho_2=\openone/2.$

We can now consider a unitary rotation around the $z$-axis 
\begin{equation}
W=e^{-i\phi\sigma_z/2},
\end{equation}
where  $\phi$ is real.  Based on elementary geometrical considerations, we can always find an angle $\phi$ such that
\begin{equation}
H_3=W H_2W^\dagger
\end{equation}
has only real elements or, equivalently, ${\rm Tr}(H_3 \sigma_y)=0.$ Clearly, $\varrho_2$ does not change under such a unitary, hence
\begin{equation}
\varrho_3=W \varrho_2 W^\dagger \equiv D.
\end{equation}
Thus, the unitary we look for is 
\begin{equation}
U=WV^\dagger.
\end{equation}
This finishes our proof. $\qed$

Using \LEMMA{lemma:pt}, we can formulate our first main result.

\DEFTHM{thm:dpteqgmpcd2N1}For single-qubit states $\varrho$ and $\sigma$ and any Hermitian operator $H_1$ 
\begin{equation}
D_{\rm DPT}(\varrho,\sigma)^2=D_{\rm GMPC}(\varrho,\sigma)^2\label{eq:eqDPTGMPC}
\end{equation}
holds when the cost function contains a single operator, $N=1.$

{\it Proof.} Based on \LEMMA{lemma:pt}, let us consider the unitary $U$ that rotates both $\varrho$ and $H_1$ into matrices with real elements only. Note that $\sigma$ is not required to have real elements. Then,  
\begin{align}
\begin{split}
D_{\rm DPT}^{\{H_1\}}(\varrho,\sigma)^2&=D_{\rm DPT}^{\{U H_1 U^\dagger \}}(U \varrho  U^{\dagger},U \sigma U^{\dagger})^2\\
&=D_{\rm GMPC}^{\{U H_1 U^\dagger \}}(U \varrho  U^{\dagger}, U \sigma U^\dagger)^2\\
&=D_{\rm GMPC}^{\{ H_1 \}}(\varrho,\sigma)^2.\label{eq:DD}
\end{split}
\end{align}
We consider the case of a single operator, $N=1.$ In \EQ{eq:DD},  in the superscript, we gave explicitly the $H_1$ operator used for the distances.  

The first equality in \EQ{eq:DD} follows from the unitary covariance of $D_{\rm DPT}.$ Indeed, under the simultaneous transformations $H_1\to UH_1U^\dagger$, $\varrho\to U\varrho U^\dagger,$ and $\sigma\to U\sigma U^\dagger,$ an admissible coupling $\varrho_{12}$ is mapped to $(U^*\otimes U)\varrho_{12}(U^T\otimes U^\dagger),$ whose marginals are $(U\varrho U^\dagger)^T$ and $U\sigma U^\dagger,$ while the corresponding cost is unchanged. The second equality in \EQ{eq:DD} can be proven using \DEFINITIONS{def:DGMCP} and \ref{def:D}, see \EQ{eq:GMPCDPT2}. The third equality in \EQ{eq:DD} follows from the unitary covariance of $D_{\rm GMPC}$ under simultaneous unitary conjugation of $H_1,\varrho,$ and $\sigma.$ This finishes our proof. $\qed$

Next, we make a statement that is valid even for larger dimensions and for an arbitrary number of operators $H_n.$

\DEFTHM{thm:sepfd}For states $\varrho$ on a finite-dimensional Hilbert space of dimension $d\ge2$ and any $N,$ the squared self-distances satisfy the inequality
\begin{equation}
D_{\rm GMPC}(\varrho,\varrho)^2\ge D_{\rm DPT}(\varrho,\varrho)^2.\label{eq:eqDPTGMPCself}
\end{equation}

{\it Proof.}  For two-dimensional systems (qubits) and for any $N\ge1$, we can bound the square of the distance as 
\begin{align}
\begin{split}
D^{\{H_1,H_2,...,H_N\}}_{\rm GMPC}(\varrho,\sigma)^2&\ge \sum_{n=1}^N D^{\{H_n\}}_{\rm GMPC}(\varrho,\sigma)^2\\
&=\sum_{n=1}^N D^{\{H_n\}}_{\rm DPT}(\varrho,\sigma)^2. \label{eq:ineq}
\end{split}
\end{align}
Based on \EQ{eq:GMPC_distance}, the inequality in \EQ{eq:ineq} holds since the minimum of a sum is not smaller than the sum of the separate minima.
The equality is  based on \THM{thm:dpteqgmpcd2N1}. Then, we can write that for qubits
\begin{align}
\begin{split}
&D^{\{H_1,H_2,...,H_N\}}_{\rm GMPC}(\varrho,\varrho)^2\ge\sum_{n=1}^N D^{\{H_n\}}_{\rm DPT}(\varrho,\varrho)^2\\
&\quad\quad=\sum_{n=1}^N I^{\rm WY}_\varrho(H_n)= D^{\{H_1,H_2,...,H_N\}}_{\rm DPT}(\varrho,\varrho)^2
\end{split}
\end{align}
holds. The inequality is based on \EQ{eq:ineq}. Both equalities are based on \EQ{eq:DrhoI}. Thus, for qubits, the square of the self-distance of $D_{\rm GMPC}$ is bounded from below by the square of the self-distance of $D_{\rm DPT}.$ 

Let us consider the $d>2$ case. We use 
\begin{align}
\begin{split}
D^{\{H_1,H_2,...,H_N\}}_{\rm GMPC}(\varrho,\sigma)^2&\ge \sum_{n=1}^N D^{\{H_n\}}_{\rm GMPC}(\varrho,\sigma)^2.
\end{split}\label{eq:GMPCN}
\end{align}
The inequality is based on the definition given in \EQ{eq:GMPC_distance}. 

Then, we need to show that 
\begin{align}
\begin{split}
&D^{\{H_n\}}_{\rm GMPC}(\varrho,\varrho)^2 \ge I^{\rm WY}_{\varrho}(H_n)\label{eq:poz}
\end{split}
\end{align}
holds. In order to prove the inequality given in \EQ{eq:poz}, let us choose as basis the eigenbasis of $H_n,$ for which
\begin{equation}
H_n^T=H_n\label{eq:HTH}
\end{equation}
holds. Then, using \EQS{eq:DrhoI}, \eqref{eq:dptdiv} and \eqref{eq:GMPCDPT}, we obtain 
\begin{align}
\begin{split}
&D_{\rm GMPC}^{\{H_n\}}(\varrho,\varrho)^2 = D_{\rm DPT}^{\{H_n\}}(\varrho^T,\varrho)^2\\
&\qquad\ge \frac 1 2 \left[I^{\rm WY}_\varrho(H_n)+I^{\rm WY}_{\varrho^T}(H_n)\right] \\
&\qquad = \frac 1 2 \left[I^{\rm WY}_\varrho(H_n)+I^{\rm WY}_{\varrho^T}(H_n^T)\right]\\
&\qquad =I^{\rm WY}_\varrho(H_n),
\end{split}
\end{align}
where the second equality follows from \EQ{eq:HTH},  while in the last equality we used
\begin{equation}
I^{\rm WY}_{\varrho^T}(H_n^T)=I^{\rm WY}_\varrho(H_n),
\end{equation}
which follows directly from \EQ{eq:WY}. Hence, the inequality given in \EQ{eq:poz} follows. Note that since both sides of \EQ{eq:poz} are invariant under simultaneous unitary transformations of $H_n$ and $\varrho,$ the inequality in \EQ{eq:poz} holds in an arbitrary basis.

Finally, it follows that 
\begin{align}
\begin{split}
&D^{\{H_1,H_2,...,H_N\}}_{\rm GMPC}(\varrho,\varrho)^2\\
&\quad\quad\ge \sum_{n=1}^N I^{\rm WY}_{\varrho}(H_n)=D^{\{H_1,H_2,...,H_N\}}_{\rm DPT}(\varrho,\varrho)^2.
\end{split}
\end{align}
This finishes our proof. $\qed$

Note  that Lafleche also investigates the self-distance of the Golse-Mouhot-Paul-Caglioti quantum Wasserstein distance and derives upper bounds in terms of the Wigner-Yanase skew information \cite{lafleche2023quantumoptimaltransportweak} .

\section{Consequences for calculating the energy minimum for bipartite systems}
\label{sec:cons}

Let us point out one of the consequences of our main results.

Based on \DEFINITION{def:D} and \EQ{eq:DrhoI}, we can relate the minimal energy of a bipartite system to the Wigner-Yanase skew information as 
\begin{align}
\begin{split}
\min_{ \varrho_{12} }\;&
\trace\left[-\sum_{n=1}^N(H_n^T \otimes H_n)  \varrho_{12} \right]=\sum_{n=1}^N \left[I^{\rm WY}_{\varrho}(H_n)-\ex{H_n^2}_\varrho\right],\\
\textrm{s.t. }&
\varrho_{12}\in\mathcal D,\\
& {\rm Tr}_2(\varrho_{12})=\varrho^T,\\
& {\rm Tr}_1(\varrho_{12})=\varrho.\label{eq:distance2}
\end{split}
\end{align}
The left-hand side is the minimal energy for a quantum state with given marginals.

Due to \THM{thm:sepfd}, we can also obtain that 
\begin{align}
\begin{split}
\min_{ \varrho_{12} }\;&
\trace\left[-\sum_{n=1}^N(H_n \otimes H_n)  \varrho_{12} \right]\ge\sum_{n=1}^N \left[I^{\rm WY}_{\varrho}(H_n)-\ex{H_n^2}_\varrho\right],\\
\textrm{s.t. }&
\varrho_{12}\in\mathcal D,\\
& {\rm Tr}_2(\varrho_{12})=\varrho,\\
& {\rm Tr}_1(\varrho_{12})=\varrho,\label{eq:distance3}
\end{split}
\end{align}
holds. Using \THM{thm:dpteqgmpcd2N1}, equality holds in \EQ{eq:distance3} if we have qubits and we have a single operator, i.e., $N=1$. Thus, we obtain a lower bound on the energy minimum of the  model in \EQ{eq:distance3} with given marginals, in terms of the  Wigner-Yanase skew information \cite{toth2026general}. Note that transposes do not appear in the formula in \EQ{eq:distance3}. 

\section{Conclusions}

We showed that the quantum Wasserstein distances defined by Golse, Mouhot, Paul, and Caglioti, and by De Palma and Trevisan are equal for qubits when a single operator appears in the cost function. We also compared their self-distances for finite-dimensional systems of dimension $d\ge2$ and an arbitrary number of operators. We found that the Golse-Mouhot-Paul-Caglioti self-distance is bounded from below by the De Palma-Trevisan self-distance.

\begin{acknowledgments}
We thank A. Ac\'{\i}n, I.~Apellaniz, G.~De Palma, M.~Eckstein, F. Fr\"owis, I.~L.~Egusquiza, C.~Klempt, J.~Ko\l ody\'nski, L.~Lafleche, M.~Mosonyi, G.~Muga, J.~Siewert, Sz.~Szalay, T.~Titkos, K. \.Zyczkowski, T.~V\'ertesi, G. Vitagliano, and D. Virosztek  for discussions. We acknowledge the support of the  EU (QuantERA MENTA, QuantERA QuSiED, COST Action CA23115), the Spanish MCIU (Grant No.~PCI2022-132947), the Basque Government (Grant No. IT1887-26), and the National Research, Development and Innovation Office of Hungary (NKFIH) (Grant No. 2019-2.1.7-ERA-NET-2021-00036, Advanced Grant No. 152794). We thank the National Research, Development and Innovation Office of Hungary (NKFIH) within the Quantum Information National Laboratory of Hungary.   We acknowledge the support of the Grant No.~PID2021-126273NB-I00 and PID2025-171880NB-I00 funded by MCIN/AEI/10.13039/501100011033 and by ``ERDF A way of making Europe''.  We thank the ``Frontline'' Research Excellence Programme of the NKFIH (Grant No. KKP133827). We thank Project no. TKP2021-NVA-04, which has been implemented with the support provided by the Ministry of Innovation and Technology of Hungary from the National Research, Development and Innovation Fund, financed under the TKP2021-NVA funding scheme.
\end{acknowledgments}

\bibliography{Bibliography2}

\begin{thebibliography}{50}%
\makeatletter
\providecommand \@ifxundefined [1]{%
 \@ifx{#1\undefined}
}%
\providecommand \@ifnum [1]{%
 \ifnum #1\expandafter \@firstoftwo
 \else \expandafter \@secondoftwo
 \fi
}%
\providecommand \@ifx [1]{%
 \ifx #1\expandafter \@firstoftwo
 \else \expandafter \@secondoftwo
 \fi
}%
\providecommand \natexlab [1]{#1}%
\providecommand \enquote  [1]{``#1''}%
\providecommand \bibnamefont  [1]{#1}%
\providecommand \bibfnamefont [1]{#1}%
\providecommand \citenamefont [1]{#1}%
\providecommand \href@noop [0]{\@secondoftwo}%
\providecommand \href [0]{\begingroup \@sanitize@url \@href}%
\providecommand \@href[1]{\@@startlink{#1}\@@href}%
\providecommand \@@href[1]{\endgroup#1\@@endlink}%
\providecommand \@sanitize@url [0]{\catcode `\\12\catcode `\$12\catcode
  `\&12\catcode `\#12\catcode `\^12\catcode `\_12\catcode `\%12\relax}%
\providecommand \@@startlink[1]{}%
\providecommand \@@endlink[0]{}%
\providecommand \url  [0]{\begingroup\@sanitize@url \@url }%
\providecommand \@url [1]{\endgroup\@href {#1}{\urlprefix }}%
\providecommand \urlprefix  [0]{URL }%
\providecommand \Eprint [0]{\href }%
\providecommand \doibase [0]{https://doi.org/}%
\providecommand \selectlanguage [0]{\@gobble}%
\providecommand \bibinfo  [0]{\@secondoftwo}%
\providecommand \bibfield  [0]{\@secondoftwo}%
\providecommand \translation [1]{[#1]}%
\providecommand \BibitemOpen [0]{}%
\providecommand \bibitemStop [0]{}%
\providecommand \bibitemNoStop [0]{.\EOS\space}%
\providecommand \EOS [0]{\spacefactor3000\relax}%
\providecommand \BibitemShut  [1]{\csname bibitem#1\endcsname}%
\let\auto@bib@innerbib\@empty
\bibitem [{\citenamefont {\.Zyczkowski}\ and\ \citenamefont
  {S{\l}omczy\'nski}(1998)}]{Zyczkowski1998TheMonge}%
  \BibitemOpen
  \bibfield  {author} {\bibinfo {author} {\bibfnamefont {K.}~\bibnamefont
  {\.Zyczkowski}}\ and\ \bibinfo {author} {\bibfnamefont {W.}~\bibnamefont
  {S{\l}omczy\'nski}},\ }\bibfield  {title} {\bibinfo {title} {The {Monge}
  distance between quantum states},\ }\href
  {https://doi.org/10.1088/0305-4470/31/45/009} {\bibfield  {journal} {\bibinfo
   {journal} {J. Phys. A: Math. Gen.}\ }\textbf {\bibinfo {volume} {31}},\
  \bibinfo {pages} {9095} (\bibinfo {year} {1998})}\BibitemShut {NoStop}%
\bibitem [{\citenamefont {\.Zyczkowski}\ and\ \citenamefont
  {S{\l}omczy\'nski}(2001)}]{Zyczkowski2001TheMonge}%
  \BibitemOpen
  \bibfield  {author} {\bibinfo {author} {\bibfnamefont {K.}~\bibnamefont
  {\.Zyczkowski}}\ and\ \bibinfo {author} {\bibfnamefont {W.}~\bibnamefont
  {S{\l}omczy\'nski}},\ }\bibfield  {title} {\bibinfo {title} {The {Monge}
  metric on the sphere and geometry of quantum states},\ }\href
  {https://doi.org/10.1088/0305-4470/34/34/311} {\bibfield  {journal} {\bibinfo
   {journal} {J. Phys. A: Math. Gen.}\ }\textbf {\bibinfo {volume} {34}},\
  \bibinfo {pages} {6689} (\bibinfo {year} {2001})}\BibitemShut {NoStop}%
\bibitem [{\citenamefont {Bengtsson}\ and\ \citenamefont
  {\.Zyczkowski}(2006)}]{Bengtsson2006Geometry}%
  \BibitemOpen
  \bibfield  {author} {\bibinfo {author} {\bibfnamefont {I.}~\bibnamefont
  {Bengtsson}}\ and\ \bibinfo {author} {\bibfnamefont {K.}~\bibnamefont
  {\.Zyczkowski}},\ }\href {https://doi.org/10.1017/CBO9780511535048} {\emph
  {\bibinfo {title} {Geometry of Quantum States: An Introduction to Quantum
  Entanglement}}}\ (\bibinfo  {publisher} {Cambridge University Press},\
  \bibinfo {year} {2006})\BibitemShut {NoStop}%
\bibitem [{\citenamefont {{P. Biane and D. Voiculescu}}(2001)}]{Biane2011Free}%
  \BibitemOpen
  \bibfield  {author} {\bibinfo {author} {\bibnamefont {{P. Biane and D.
  Voiculescu}}},\ }\bibfield  {title} {\bibinfo {title} {{A free probability
  analogue of the Wasserstein metric on the trace-state space}},\ }\href
  {https://doi.org/10.1007/s00039-001-8226-4} {\bibfield  {journal} {\bibinfo
  {journal} {GAFA, Geom. Funct. Anal.}\ }\textbf {\bibinfo {volume} {11}},\
  \bibinfo {pages} {1125} (\bibinfo {year} {2001})}\BibitemShut {NoStop}%
\bibitem [{\citenamefont {Carlen}\ and\ \citenamefont
  {Maas}(2014)}]{CarlenMaas2014Analog}%
  \BibitemOpen
  \bibfield  {author} {\bibinfo {author} {\bibfnamefont {E.}~\bibnamefont
  {Carlen}}\ and\ \bibinfo {author} {\bibfnamefont {J.}~\bibnamefont {Maas}},\
  }\bibfield  {title} {\bibinfo {title} {{An Analog of the 2-Wasserstein Metric
  in Non-Commutative Probability Under Which the Fermionic Fokker-Planck
  Equation is Gradient Flow for the Entropy}},\ }\href
  {https://doi.org/10.1007/s00220-014-2124-8} {\bibfield  {journal} {\bibinfo
  {journal} {Commun. Math. Phys.}\ }\textbf {\bibinfo {volume} {331}},\
  \bibinfo {pages} {887} (\bibinfo {year} {2014})}\BibitemShut {NoStop}%
\bibitem [{\citenamefont {Carlen}\ and\ \citenamefont
  {Maas}(2017)}]{CarlenMaas2017Gradient}%
  \BibitemOpen
  \bibfield  {author} {\bibinfo {author} {\bibfnamefont {E.~A.}\ \bibnamefont
  {Carlen}}\ and\ \bibinfo {author} {\bibfnamefont {J.}~\bibnamefont {Maas}},\
  }\bibfield  {title} {\bibinfo {title} {{Gradient flow and entropy
  inequalities for quantum Markov semigroups with detailed balance}},\ }\href
  {https://doi.org/10.1016/j.jfa.2017.05.003} {\bibfield  {journal} {\bibinfo
  {journal} {J. Funct. Anal.}\ }\textbf {\bibinfo {volume} {273}},\ \bibinfo
  {pages} {1810} (\bibinfo {year} {2017})}\BibitemShut {NoStop}%
\bibitem [{\citenamefont {Carlen}\ and\ \citenamefont
  {Maas}(2020)}]{CarlenMaas2020Non-commutative}%
  \BibitemOpen
  \bibfield  {author} {\bibinfo {author} {\bibfnamefont {E.~A.}\ \bibnamefont
  {Carlen}}\ and\ \bibinfo {author} {\bibfnamefont {J.}~\bibnamefont {Maas}},\
  }\bibfield  {title} {\bibinfo {title} {Non-commutative calculus, optimal
  transport and functional inequalities in dissipative quantum systems},\
  }\href {https://doi.org/10.1007/s10955-019-02434-w} {\bibfield  {journal}
  {\bibinfo  {journal} {J. Stat. Phys.}\ }\textbf {\bibinfo {volume} {178}},\
  \bibinfo {pages} {319} (\bibinfo {year} {2020})}\BibitemShut {NoStop}%
\bibitem [{\citenamefont {Datta}\ and\ \citenamefont
  {Rouz\'e}(2019)}]{DattaRouze2019Concentration}%
  \BibitemOpen
  \bibfield  {author} {\bibinfo {author} {\bibfnamefont {N.}~\bibnamefont
  {Datta}}\ and\ \bibinfo {author} {\bibfnamefont {C.}~\bibnamefont
  {Rouz\'e}},\ }\bibfield  {title} {\bibinfo {title} {Concentration of quantum
  states from quantum functional and transportation cost inequalities},\ }\href
  {https://doi.org/10.1063/1.5023210} {\bibfield  {journal} {\bibinfo
  {journal} {J. Math. Phys.}\ }\textbf {\bibinfo {volume} {60}},\ \bibinfo
  {pages} {012202} (\bibinfo {year} {2019})}\BibitemShut {NoStop}%
\bibitem [{\citenamefont {Datta}\ and\ \citenamefont
  {Rouz\'e}(2020)}]{DattaRouze2020Relating}%
  \BibitemOpen
  \bibfield  {author} {\bibinfo {author} {\bibfnamefont {N.}~\bibnamefont
  {Datta}}\ and\ \bibinfo {author} {\bibfnamefont {C.}~\bibnamefont
  {Rouz\'e}},\ }\bibfield  {title} {\bibinfo {title} {{Relating relative
  entropy, optimal transport and Fisher information: A quantum HWI
  inequality}},\ }\href {https://doi.org/10.1007/s00023-020-00891-8} {\bibfield
   {journal} {\bibinfo  {journal} {Ann. Henri Poincar\'e}\ }\textbf {\bibinfo
  {volume} {21}},\ \bibinfo {pages} {2115} (\bibinfo {year}
  {2020})}\BibitemShut {NoStop}%
\bibitem [{\citenamefont {Golse}\ \emph {et~al.}(2016)\citenamefont {Golse},
  \citenamefont {Mouhot},\ and\ \citenamefont {Paul}}]{Golse2016On}%
  \BibitemOpen
  \bibfield  {author} {\bibinfo {author} {\bibfnamefont {F.}~\bibnamefont
  {Golse}}, \bibinfo {author} {\bibfnamefont {C.}~\bibnamefont {Mouhot}},\ and\
  \bibinfo {author} {\bibfnamefont {T.}~\bibnamefont {Paul}},\ }\bibfield
  {title} {\bibinfo {title} {On the mean field and classical limits of quantum
  mechanics},\ }\href {https://doi.org/10.1007/s00220-015-2485-7} {\bibfield
  {journal} {\bibinfo  {journal} {Commun. Math. Phys.}\ }\textbf {\bibinfo
  {volume} {343}},\ \bibinfo {pages} {165} (\bibinfo {year}
  {2016})}\BibitemShut {NoStop}%
\bibitem [{\citenamefont {Golse}\ and\ \citenamefont
  {Paul}(2017)}]{Golse2017The}%
  \BibitemOpen
  \bibfield  {author} {\bibinfo {author} {\bibfnamefont {F.}~\bibnamefont
  {Golse}}\ and\ \bibinfo {author} {\bibfnamefont {T.}~\bibnamefont {Paul}},\
  }\bibfield  {title} {\bibinfo {title} {The {Schr{\"o}dinger} equation in the
  mean-field and semiclassical regime},\ }\href
  {https://doi.org/10.1007/s00205-016-1031-x} {\bibfield  {journal} {\bibinfo
  {journal} {Arch. Ration. Mech. Anal.}\ }\textbf {\bibinfo {volume} {223}},\
  \bibinfo {pages} {57} (\bibinfo {year} {2017})}\BibitemShut {NoStop}%
\bibitem [{\citenamefont {Golse}\ and\ \citenamefont
  {Paul}(2018)}]{Golse2018Wave}%
  \BibitemOpen
  \bibfield  {author} {\bibinfo {author} {\bibfnamefont {F.}~\bibnamefont
  {Golse}}\ and\ \bibinfo {author} {\bibfnamefont {T.}~\bibnamefont {Paul}},\
  }\bibfield  {title} {\bibinfo {title} {Wave packets and the quadratic
  {Monge}-{Kantorovich} distance in quantum mechanics},\ }\href
  {https://doi.org/https://doi.org/10.1016/j.crma.2017.12.007} {\bibfield
  {journal} {\bibinfo  {journal} {Comptes Rendus Math.}\ }\textbf {\bibinfo
  {volume} {356}},\ \bibinfo {pages} {177} (\bibinfo {year}
  {2018})}\BibitemShut {NoStop}%
\bibitem [{\citenamefont {Golse}(2018)}]{Golse2018TheQuantum}%
  \BibitemOpen
  \bibfield  {author} {\bibinfo {author} {\bibfnamefont {F.}~\bibnamefont
  {Golse}},\ }\bibfield  {title} {\bibinfo {title} {The quantum {$N$}-body
  problem in the mean-field and semiclassical regime},\ }\href
  {https://doi.org/10.1098/rsta.2017.0229} {\bibfield  {journal} {\bibinfo
  {journal} {Phil. Trans. R. Soc. A}\ }\textbf {\bibinfo {volume} {376}},\
  \bibinfo {pages} {20170229} (\bibinfo {year} {2018})}\BibitemShut {NoStop}%
\bibitem [{\citenamefont {Caglioti}\ \emph {et~al.}(2020)\citenamefont
  {Caglioti}, \citenamefont {Golse},\ and\ \citenamefont
  {Paul}}]{Caglioti2020Quantum}%
  \BibitemOpen
  \bibfield  {author} {\bibinfo {author} {\bibfnamefont {E.}~\bibnamefont
  {Caglioti}}, \bibinfo {author} {\bibfnamefont {F.}~\bibnamefont {Golse}},\
  and\ \bibinfo {author} {\bibfnamefont {T.}~\bibnamefont {Paul}},\ }\bibfield
  {title} {\bibinfo {title} {Quantum optimal transport is cheaper},\ }\href
  {https://doi.org/10.1007/s10955-020-02571-7} {\bibfield  {journal} {\bibinfo
  {journal} {J. Stat. Phys.}\ }\textbf {\bibinfo {volume} {181}},\ \bibinfo
  {pages} {149} (\bibinfo {year} {2020})}\BibitemShut {NoStop}%
\bibitem [{\citenamefont {Caglioti}\ \emph {et~al.}(2023)\citenamefont
  {Caglioti}, \citenamefont {Golse},\ and\ \citenamefont
  {Paul}}]{Caglioti2023Towards}%
  \BibitemOpen
  \bibfield  {author} {\bibinfo {author} {\bibfnamefont {E.}~\bibnamefont
  {Caglioti}}, \bibinfo {author} {\bibfnamefont {F.}~\bibnamefont {Golse}},\
  and\ \bibinfo {author} {\bibfnamefont {T.}~\bibnamefont {Paul}},\ }\bibfield
  {title} {\bibinfo {title} {Towards optimal transport for quantum densities},\
  }\href {https://doi.org/10.2422/2036-2145.202106_011} {\bibfield  {journal}
  {\bibinfo  {journal} {Annali Scuola Normale Superiore - Classe di Scienze}\
  }\textbf {\bibinfo {volume} {24}},\ \bibinfo {pages} {1981} (\bibinfo {year}
  {2023})}\BibitemShut {NoStop}%
\bibitem [{\citenamefont
  {Lafleche}(2023)}]{lafleche2023quantumoptimaltransportweak}%
  \BibitemOpen
  \bibfield  {author} {\bibinfo {author} {\bibfnamefont {L.}~\bibnamefont
  {Lafleche}},\ }\bibfield  {title} {\bibinfo {title} {Quantum optimal
  transport and weak topologies},\ }\href {https://arxiv.org/abs/2306.12944}
  {\bibfield  {journal} {\bibinfo  {journal} {arXiv:2306.12944}\ } (\bibinfo
  {year} {2023})}\BibitemShut {NoStop}%
\bibitem [{\citenamefont {De~Palma}\ and\ \citenamefont
  {Trevisan}(2021)}]{DePalma2021Quantum}%
  \BibitemOpen
  \bibfield  {author} {\bibinfo {author} {\bibfnamefont {G.}~\bibnamefont
  {De~Palma}}\ and\ \bibinfo {author} {\bibfnamefont {D.}~\bibnamefont
  {Trevisan}},\ }\bibfield  {title} {\bibinfo {title} {Quantum optimal
  transport with quantum channels},\ }\href
  {https://doi.org/10.1007/s00023-021-01042-3} {\bibfield  {journal} {\bibinfo
  {journal} {Ann. Henri Poincar{\'e}}\ }\textbf {\bibinfo {volume} {22}},\
  \bibinfo {pages} {3199} (\bibinfo {year} {2021})}\BibitemShut {NoStop}%
\bibitem [{\citenamefont {De~Palma}\ and\ \citenamefont
  {Trevisan}(2024)}]{DePalma2024QuantumOptimal}%
  \BibitemOpen
  \bibfield  {author} {\bibinfo {author} {\bibfnamefont {G.}~\bibnamefont
  {De~Palma}}\ and\ \bibinfo {author} {\bibfnamefont {D.}~\bibnamefont
  {Trevisan}},\ }\bibinfo {title} {Quantum optimal transport: Quantum channels
  and qubits},\ in\ \href {https://doi.org/10.1007/978-3-031-50466-2_4} {\emph
  {\bibinfo {booktitle} {Optimal Transport on Quantum Structures}}},\ \bibinfo
  {editor} {edited by\ \bibinfo {editor} {\bibfnamefont {J.}~\bibnamefont
  {Maas}}, \bibinfo {editor} {\bibfnamefont {S.}~\bibnamefont {Rademacher}},
  \bibinfo {editor} {\bibfnamefont {T.}~\bibnamefont {Titkos}},\ and\ \bibinfo
  {editor} {\bibfnamefont {D.}~\bibnamefont {Virosztek}}}\ (\bibinfo
  {publisher} {Springer Nature Switzerland},\ \bibinfo {address} {Cham},\
  \bibinfo {year} {2024})\ pp.\ \bibinfo {pages} {203--239}\BibitemShut
  {NoStop}%
\bibitem [{\citenamefont {De~Palma}\ \emph {et~al.}(2021)\citenamefont
  {De~Palma}, \citenamefont {Marvian}, \citenamefont {Trevisan},\ and\
  \citenamefont {Lloyd}}]{DePalma2021TheQuantum}%
  \BibitemOpen
  \bibfield  {author} {\bibinfo {author} {\bibfnamefont {G.}~\bibnamefont
  {De~Palma}}, \bibinfo {author} {\bibfnamefont {M.}~\bibnamefont {Marvian}},
  \bibinfo {author} {\bibfnamefont {D.}~\bibnamefont {Trevisan}},\ and\
  \bibinfo {author} {\bibfnamefont {S.}~\bibnamefont {Lloyd}},\ }\bibfield
  {title} {\bibinfo {title} {The quantum {Wasserstein} distance of order 1},\
  }\href {https://doi.org/10.1109/TIT.2021.3076442} {\bibfield  {journal}
  {\bibinfo  {journal} {IEEE Trans. Inf. Theory}\ }\textbf {\bibinfo {volume}
  {67}},\ \bibinfo {pages} {6627} (\bibinfo {year} {2021})}\BibitemShut
  {NoStop}%
\bibitem [{\citenamefont {{S. Friedland, M. Eckstein, S. Cole, and K.
  \.Zyczkowski}}(2022)}]{Friedland2022Quantum}%
  \BibitemOpen
  \bibfield  {author} {\bibinfo {author} {\bibnamefont {{S. Friedland, M.
  Eckstein, S. Cole, and K. \.Zyczkowski}}},\ }\bibfield  {title} {\bibinfo
  {title} {{Quantum Monge--Kantorovich problem and transport distance between
  density matrices}},\ }\href {https://doi.org/10.1103/PhysRevLett.129.110402}
  {\bibfield  {journal} {\bibinfo  {journal} {{Phys. Rev. Lett.}}\ }\textbf
  {\bibinfo {volume} {129}},\ \bibinfo {pages} {110402} (\bibinfo {year}
  {2022})}\BibitemShut {NoStop}%
\bibitem [{\citenamefont {Cole}\ \emph {et~al.}(2023)\citenamefont {Cole},
  \citenamefont {Eckstein}, \citenamefont {Friedland},\ and\ \citenamefont
  {{\.{Z}}yczkowski}}]{Cole2023OnQuantum}%
  \BibitemOpen
  \bibfield  {author} {\bibinfo {author} {\bibfnamefont {S.}~\bibnamefont
  {Cole}}, \bibinfo {author} {\bibfnamefont {M.}~\bibnamefont {Eckstein}},
  \bibinfo {author} {\bibfnamefont {S.}~\bibnamefont {Friedland}},\ and\
  \bibinfo {author} {\bibfnamefont {K.}~\bibnamefont {{\.{Z}}yczkowski}},\
  }\bibfield  {title} {\bibinfo {title} {On quantum optimal transport},\ }\href
  {https://doi.org/10.1007/s11040-023-09456-7} {\bibfield  {journal} {\bibinfo
  {journal} {Mathematical Physics, Analysis and Geometry}\ }\textbf {\bibinfo
  {volume} {26}},\ \bibinfo {pages} {14} (\bibinfo {year} {2023})}\BibitemShut
  {NoStop}%
\bibitem [{\citenamefont {Bistro\'n}\ \emph {et~al.}(2023)\citenamefont
  {Bistro\'n}, \citenamefont {Eckstein},\ and\ \citenamefont
  {\.Zyczkowski}}]{Bistron2023Monotonicity}%
  \BibitemOpen
  \bibfield  {author} {\bibinfo {author} {\bibfnamefont {R.}~\bibnamefont
  {Bistro\'n}}, \bibinfo {author} {\bibfnamefont {M.}~\bibnamefont
  {Eckstein}},\ and\ \bibinfo {author} {\bibfnamefont {K.}~\bibnamefont
  {\.Zyczkowski}},\ }\bibfield  {title} {\bibinfo {title} {{Monotonicity of a
  quantum 2-Wasserstein distance}},\ }\href
  {https://doi.org/10.1088/1751-8121/acb9c8} {\bibfield  {journal} {\bibinfo
  {journal} {J. Phys. A: Math. Theor.}\ }\textbf {\bibinfo {volume} {56}},\
  \bibinfo {pages} {095301} (\bibinfo {year} {2023})}\BibitemShut {NoStop}%
\bibitem [{\citenamefont {Geh\'er}\ \emph {et~al.}(2023)\citenamefont
  {Geh\'er}, \citenamefont {Pitrik}, \citenamefont {Titkos},\ and\
  \citenamefont {Virosztek}}]{Geher2023Quantum}%
  \BibitemOpen
  \bibfield  {author} {\bibinfo {author} {\bibfnamefont {G.~P.}\ \bibnamefont
  {Geh\'er}}, \bibinfo {author} {\bibfnamefont {J.}~\bibnamefont {Pitrik}},
  \bibinfo {author} {\bibfnamefont {T.}~\bibnamefont {Titkos}},\ and\ \bibinfo
  {author} {\bibfnamefont {D.}~\bibnamefont {Virosztek}},\ }\bibfield  {title}
  {\bibinfo {title} {{Quantum Wasserstein isometries on the qubit state
  space}},\ }\href {https://doi.org/https://doi.org/10.1016/j.jmaa.2022.126955}
  {\bibfield  {journal} {\bibinfo  {journal} {J. Math. Anal. Appl.}\ }\textbf
  {\bibinfo {volume} {522}},\ \bibinfo {pages} {126955} (\bibinfo {year}
  {2023})}\BibitemShut {NoStop}%
\bibitem [{\citenamefont {Li}\ \emph {et~al.}(2025)\citenamefont {Li},
  \citenamefont {Bu}, \citenamefont {Enshan~Koh}, \citenamefont {Jaffe},\ and\
  \citenamefont {Lloyd}}]{Li2025Wasserstein}%
  \BibitemOpen
  \bibfield  {author} {\bibinfo {author} {\bibfnamefont {L.}~\bibnamefont
  {Li}}, \bibinfo {author} {\bibfnamefont {K.}~\bibnamefont {Bu}}, \bibinfo
  {author} {\bibfnamefont {D.}~\bibnamefont {Enshan~Koh}}, \bibinfo {author}
  {\bibfnamefont {A.}~\bibnamefont {Jaffe}},\ and\ \bibinfo {author}
  {\bibfnamefont {S.}~\bibnamefont {Lloyd}},\ }\bibfield  {title} {\bibinfo
  {title} {Wasserstein complexity of quantum circuits},\ }\href
  {https://doi.org/10.1088/1751-8121/ade381} {\bibfield  {journal} {\bibinfo
  {journal} {J. Phys. A: Math. Theor.}\ }\textbf {\bibinfo {volume} {58}},\
  \bibinfo {pages} {265302} (\bibinfo {year} {2025})}\BibitemShut {NoStop}%
\bibitem [{\citenamefont {Trevisan}(2025)}]{Trevisan2025Quantum}%
  \BibitemOpen
  \bibfield  {author} {\bibinfo {author} {\bibfnamefont {D.}~\bibnamefont
  {Trevisan}},\ }\bibfield  {title} {\bibinfo {title} {Quantum optimal
  transport: an invitation},\ }\href
  {https://doi.org/10.1007/s40574-024-00428-5} {\bibfield  {journal} {\bibinfo
  {journal} {Bollettino dell'Unione Matematica Italiana}\ }\textbf {\bibinfo
  {volume} {18}},\ \bibinfo {pages} {347} (\bibinfo {year} {2025})}\BibitemShut
  {NoStop}%
\bibitem [{\citenamefont {Beatty}(2026)}]{Beatty2026Wasserstein}%
  \BibitemOpen
  \bibfield  {author} {\bibinfo {author} {\bibfnamefont {E.}~\bibnamefont
  {Beatty}},\ }\bibfield  {title} {\bibinfo {title} {Wasserstein distances on
  quantum structures: An overview},\ }\bibfield  {journal} {\bibinfo  {journal}
  {Rev. Math. Phys.}\ }\href {https://doi.org/10.1142/s0129055x26300037}
  {10.1142/s0129055x26300037} (\bibinfo {year} {2026})\BibitemShut {NoStop}%
\bibitem [{\citenamefont {Kiani}\ \emph {et~al.}(2022)\citenamefont {Kiani},
  \citenamefont {{De Palma}}, \citenamefont {Marvian}, \citenamefont {Liu},\
  and\ \citenamefont {Lloyd}}]{Kiani2022Learning}%
  \BibitemOpen
  \bibfield  {author} {\bibinfo {author} {\bibfnamefont {B.~T.}\ \bibnamefont
  {Kiani}}, \bibinfo {author} {\bibfnamefont {G.}~\bibnamefont {{De Palma}}},
  \bibinfo {author} {\bibfnamefont {M.}~\bibnamefont {Marvian}}, \bibinfo
  {author} {\bibfnamefont {Z.-W.}\ \bibnamefont {Liu}},\ and\ \bibinfo {author}
  {\bibfnamefont {S.}~\bibnamefont {Lloyd}},\ }\bibfield  {title} {\bibinfo
  {title} {Learning quantum data with the quantum earth mover's distance},\
  }\href {https://doi.org/10.1088/2058-9565/ac79c9} {\bibfield  {journal}
  {\bibinfo  {journal} {Quantum Sci. Technol.}\ }\textbf {\bibinfo {volume}
  {7}},\ \bibinfo {pages} {045002} (\bibinfo {year} {2022})}\BibitemShut
  {NoStop}%
\bibitem [{\citenamefont {Malyshev}\ \emph {et~al.}(2026)\citenamefont
  {Malyshev}, \citenamefont {Linsel}, \citenamefont {Grusdt}, \citenamefont
  {Bohrdt}, \citenamefont {Demler},\ and\ \citenamefont
  {Morera}}]{malyshev2026distancelearningprojectivemeasurements}%
  \BibitemOpen
  \bibfield  {author} {\bibinfo {author} {\bibfnamefont {O.}~\bibnamefont
  {Malyshev}}, \bibinfo {author} {\bibfnamefont {S.~M.}\ \bibnamefont
  {Linsel}}, \bibinfo {author} {\bibfnamefont {F.}~\bibnamefont {Grusdt}},
  \bibinfo {author} {\bibfnamefont {A.}~\bibnamefont {Bohrdt}}, \bibinfo
  {author} {\bibfnamefont {E.}~\bibnamefont {Demler}},\ and\ \bibinfo {author}
  {\bibfnamefont {I.}~\bibnamefont {Morera}},\ }\bibfield  {title} {\bibinfo
  {title} {Distance learning from projective measurements as an
  information-geometric probe of many-body physics},\ }\href
  {https://arxiv.org/abs/2603.13485} {\bibfield  {journal} {\bibinfo  {journal}
  {arXiv:2603.13485}\ } (\bibinfo {year} {2026})}\BibitemShut {NoStop}%
\bibitem [{\citenamefont {Wigner}\ and\ \citenamefont
  {Yanase}(1963)}]{Wigner1963INFORMATION}%
  \BibitemOpen
  \bibfield  {author} {\bibinfo {author} {\bibfnamefont {E.~P.}\ \bibnamefont
  {Wigner}}\ and\ \bibinfo {author} {\bibfnamefont {M.~M.}\ \bibnamefont
  {Yanase}},\ }\bibfield  {title} {\bibinfo {title} {Information contents of
  distributions},\ }\href {https://doi.org/10.1073/pnas.49.6.910} {\bibfield
  {journal} {\bibinfo  {journal} {Proc. Natl. Acad. Sci. U.S.A.}\ }\textbf
  {\bibinfo {volume} {49}},\ \bibinfo {pages} {910} (\bibinfo {year}
  {1963})}\BibitemShut {NoStop}%
\bibitem [{\citenamefont {Camacho}\ and\ \citenamefont
  {Fauseweh}(2025)}]{Camacho2025Critical}%
  \BibitemOpen
  \bibfield  {author} {\bibinfo {author} {\bibfnamefont {G.}~\bibnamefont
  {Camacho}}\ and\ \bibinfo {author} {\bibfnamefont {B.}~\bibnamefont
  {Fauseweh}},\ }\bibfield  {title} {\bibinfo {title} {{Critical scaling of the
  quantum Wasserstein distance}},\ }\href {https://doi.org/10.1103/hbgk-4khg}
  {\bibfield  {journal} {\bibinfo  {journal} {Phys. Rev. Res.}\ }\textbf
  {\bibinfo {volume} {7}},\ \bibinfo {pages} {043223} (\bibinfo {year}
  {2025})}\BibitemShut {NoStop}%
\bibitem [{\citenamefont {Bunth}\ \emph
  {et~al.}(2025{\natexlab{a}})\citenamefont {Bunth}, \citenamefont {Pitrik},
  \citenamefont {Titkos},\ and\ \citenamefont
  {Virosztek}}]{bunth2025wassersteindistancesdivergencesorder}%
  \BibitemOpen
  \bibfield  {author} {\bibinfo {author} {\bibfnamefont {G.}~\bibnamefont
  {Bunth}}, \bibinfo {author} {\bibfnamefont {J.}~\bibnamefont {Pitrik}},
  \bibinfo {author} {\bibfnamefont {T.}~\bibnamefont {Titkos}},\ and\ \bibinfo
  {author} {\bibfnamefont {D.}~\bibnamefont {Virosztek}},\ }\bibfield  {title}
  {\bibinfo {title} {Wasserstein distances and divergences of order $p$ by
  quantum channels},\ }\href {https://arxiv.org/abs/2501.08066} {\bibfield
  {journal} {\bibinfo  {journal} {arXiv:2501.08066}\ } (\bibinfo {year}
  {2025}{\natexlab{a}})}\BibitemShut {NoStop}%
\bibitem [{\citenamefont {Bunth}\ \emph
  {et~al.}(2025{\natexlab{b}})\citenamefont {Bunth}, \citenamefont {Pitrik},
  \citenamefont {Titkos},\ and\ \citenamefont
  {Virosztek}}]{bunth2025strongkantorovichdualityquantum}%
  \BibitemOpen
  \bibfield  {author} {\bibinfo {author} {\bibfnamefont {G.}~\bibnamefont
  {Bunth}}, \bibinfo {author} {\bibfnamefont {J.}~\bibnamefont {Pitrik}},
  \bibinfo {author} {\bibfnamefont {T.}~\bibnamefont {Titkos}},\ and\ \bibinfo
  {author} {\bibfnamefont {D.}~\bibnamefont {Virosztek}},\ }\bibfield  {title}
  {\bibinfo {title} {Strong {Kantorovich} duality for quantum optimal transport
  with generic cost and optimal couplings on quantum bits},\ }\href
  {https://arxiv.org/abs/2510.26326} {\bibfield  {journal} {\bibinfo  {journal}
  {arXiv:2510.26326}\ } (\bibinfo {year} {2025}{\natexlab{b}})}\BibitemShut
  {NoStop}%
\bibitem [{\citenamefont {Bunth}\ \emph {et~al.}(2024)\citenamefont {Bunth},
  \citenamefont {Pitrik}, \citenamefont {Titkos},\ and\ \citenamefont
  {Virosztek}}]{Bunth2024MetricProperty}%
  \BibitemOpen
  \bibfield  {author} {\bibinfo {author} {\bibfnamefont {G.}~\bibnamefont
  {Bunth}}, \bibinfo {author} {\bibfnamefont {J.}~\bibnamefont {Pitrik}},
  \bibinfo {author} {\bibfnamefont {T.}~\bibnamefont {Titkos}},\ and\ \bibinfo
  {author} {\bibfnamefont {D.}~\bibnamefont {Virosztek}},\ }\bibfield  {title}
  {\bibinfo {title} {Metric property of quantum {Wasserstein} divergences},\
  }\href {https://doi.org/10.1103/PhysRevA.110.022211} {\bibfield  {journal}
  {\bibinfo  {journal} {Phys. Rev. A}\ }\textbf {\bibinfo {volume} {110}},\
  \bibinfo {pages} {022211} (\bibinfo {year} {2024})}\BibitemShut {NoStop}%
\bibitem [{\citenamefont
  {Wirth}(2025)}]{wirth2025triangleinequalityquantumwasserstein}%
  \BibitemOpen
  \bibfield  {author} {\bibinfo {author} {\bibfnamefont {M.}~\bibnamefont
  {Wirth}},\ }\bibfield  {title} {\bibinfo {title} {Triangle inequality for a
  quantum {Wasserstein} divergence},\ }\href {https://arxiv.org/abs/2511.20450}
  {\bibfield  {journal} {\bibinfo  {journal} {arXiv:2511.20450}\ } (\bibinfo
  {year} {2025})}\BibitemShut {NoStop}%
\bibitem [{\citenamefont {T{\'{o}}th}\ and\ \citenamefont
  {Pitrik}(2023)}]{Toth2023QuantumWasserstein}%
  \BibitemOpen
  \bibfield  {author} {\bibinfo {author} {\bibfnamefont {G.}~\bibnamefont
  {T{\'{o}}th}}\ and\ \bibinfo {author} {\bibfnamefont {J.}~\bibnamefont
  {Pitrik}},\ }\bibfield  {title} {\bibinfo {title} {Quantum {W}asserstein
  distance based on an optimization over separable states},\ }\href
  {https://doi.org/10.22331/q-2023-10-16-1143} {\bibfield  {journal} {\bibinfo
  {journal} {{Quantum}}\ }\textbf {\bibinfo {volume} {7}},\ \bibinfo {pages}
  {1143} (\bibinfo {year} {2023})}\BibitemShut {NoStop}%
\bibitem [{\citenamefont {Horodecki}\ \emph {et~al.}(2009)\citenamefont
  {Horodecki}, \citenamefont {Horodecki}, \citenamefont {Horodecki},\ and\
  \citenamefont {Horodecki}}]{Horodecki2009Quantum}%
  \BibitemOpen
  \bibfield  {author} {\bibinfo {author} {\bibfnamefont {R.}~\bibnamefont
  {Horodecki}}, \bibinfo {author} {\bibfnamefont {P.}~\bibnamefont
  {Horodecki}}, \bibinfo {author} {\bibfnamefont {M.}~\bibnamefont
  {Horodecki}},\ and\ \bibinfo {author} {\bibfnamefont {K.}~\bibnamefont
  {Horodecki}},\ }\bibfield  {title} {\bibinfo {title} {Quantum entanglement},\
  }\href {https://doi.org/10.1103/RevModPhys.81.865} {\bibfield  {journal}
  {\bibinfo  {journal} {Rev. Mod. Phys.}\ }\textbf {\bibinfo {volume} {81}},\
  \bibinfo {pages} {865} (\bibinfo {year} {2009})}\BibitemShut {NoStop}%
\bibitem [{\citenamefont {G{\"u}hne}\ and\ \citenamefont
  {T{\'o}th}(2009)}]{Guhne2009Entanglement}%
  \BibitemOpen
  \bibfield  {author} {\bibinfo {author} {\bibfnamefont {O.}~\bibnamefont
  {G{\"u}hne}}\ and\ \bibinfo {author} {\bibfnamefont {G.}~\bibnamefont
  {T{\'o}th}},\ }\bibfield  {title} {\bibinfo {title} {Entanglement
  detection},\ }\href
  {https://doi.org/https://doi.org/10.1016/j.physrep.2009.02.004} {\bibfield
  {journal} {\bibinfo  {journal} {Phys. Rep.}\ }\textbf {\bibinfo {volume}
  {474}},\ \bibinfo {pages} {1} (\bibinfo {year} {2009})}\BibitemShut {NoStop}%
\bibitem [{\citenamefont {Friis}\ \emph {et~al.}(2019)\citenamefont {Friis},
  \citenamefont {Vitagliano}, \citenamefont {Malik},\ and\ \citenamefont
  {Huber}}]{Friis2019}%
  \BibitemOpen
  \bibfield  {author} {\bibinfo {author} {\bibfnamefont {N.}~\bibnamefont
  {Friis}}, \bibinfo {author} {\bibfnamefont {G.}~\bibnamefont {Vitagliano}},
  \bibinfo {author} {\bibfnamefont {M.}~\bibnamefont {Malik}},\ and\ \bibinfo
  {author} {\bibfnamefont {M.}~\bibnamefont {Huber}},\ }\bibfield  {title}
  {\bibinfo {title} {Entanglement certification from theory to experiment},\
  }\href {https://doi.org/10.1038/s42254-018-0003-5} {\bibfield  {journal}
  {\bibinfo  {journal} {Nat. Rev. Phys.}\ }\textbf {\bibinfo {volume} {1}},\
  \bibinfo {pages} {72} (\bibinfo {year} {2019})}\BibitemShut {NoStop}%
\bibitem [{\citenamefont {Giovannetti}\ \emph {et~al.}(2004)\citenamefont
  {Giovannetti}, \citenamefont {Lloyd},\ and\ \citenamefont
  {Maccone}}]{Giovannetti2004Quantum-Enhanced}%
  \BibitemOpen
  \bibfield  {author} {\bibinfo {author} {\bibfnamefont {V.}~\bibnamefont
  {Giovannetti}}, \bibinfo {author} {\bibfnamefont {S.}~\bibnamefont {Lloyd}},\
  and\ \bibinfo {author} {\bibfnamefont {L.}~\bibnamefont {Maccone}},\
  }\bibfield  {title} {\bibinfo {title} {Quantum-enhanced measurements: Beating
  the standard quantum limit},\ }\href
  {https://doi.org/10.1126/science.1104149} {\bibfield  {journal} {\bibinfo
  {journal} {Science}\ }\textbf {\bibinfo {volume} {306}},\ \bibinfo {pages}
  {1330} (\bibinfo {year} {2004})}\BibitemShut {NoStop}%
\bibitem [{\citenamefont {Paris}(2009)}]{Paris2009QUANTUM}%
  \BibitemOpen
  \bibfield  {author} {\bibinfo {author} {\bibfnamefont {M.~G.~A.}\
  \bibnamefont {Paris}},\ }\bibfield  {title} {\bibinfo {title} {Quantum
  estimation for quantum technology},\ }\href
  {https://doi.org/10.1142/S0219749909004839} {\bibfield  {journal} {\bibinfo
  {journal} {Int. J. Quant. Inf.}\ }\textbf {\bibinfo {volume} {07}},\ \bibinfo
  {pages} {125} (\bibinfo {year} {2009})}\BibitemShut {NoStop}%
\bibitem [{\citenamefont {Demkowicz-Dobrza\'nski}\ \emph
  {et~al.}(2015)\citenamefont {Demkowicz-Dobrza\'nski}, \citenamefont
  {Jarzyna},\ and\ \citenamefont
  {Ko{\l}ody\'nski}}]{Demkowicz-Dobrzanski2014Quantum}%
  \BibitemOpen
  \bibfield  {author} {\bibinfo {author} {\bibfnamefont {R.}~\bibnamefont
  {Demkowicz-Dobrza\'nski}}, \bibinfo {author} {\bibfnamefont {M.}~\bibnamefont
  {Jarzyna}},\ and\ \bibinfo {author} {\bibfnamefont {J.}~\bibnamefont
  {Ko{\l}ody\'nski}},\ }\bibfield  {title} {\bibinfo {title} {Chapter four -
  {Quantum} limits in optical interferometry},\ }\href
  {https://doi.org/10.1016/bs.po.2015.02.003} {\bibfield  {journal} {\bibinfo
  {journal} {Prog. Optics}\ }\textbf {\bibinfo {volume} {60}},\ \bibinfo
  {pages} {345 } (\bibinfo {year} {2015})},\ \Eprint
  {https://arxiv.org/abs/arXiv:1405.7703} {arXiv:1405.7703} \BibitemShut
  {NoStop}%
\bibitem [{\citenamefont {Pezze}\ and\ \citenamefont
  {Smerzi}(2014)}]{Pezze2014Quantum}%
  \BibitemOpen
  \bibfield  {author} {\bibinfo {author} {\bibfnamefont {L.}~\bibnamefont
  {Pezze}}\ and\ \bibinfo {author} {\bibfnamefont {A.}~\bibnamefont {Smerzi}},\
  }\bibfield  {title} {\bibinfo {title} {Quantum theory of phase estimation},\
  }in\ \href@noop {} {\emph {\bibinfo {booktitle} {Atom Interferometry (Proc.
  Int. School of Physics 'Enrico Fermi', Course 188, Varenna)}}},\ \bibinfo
  {editor} {edited by\ \bibinfo {editor} {\bibfnamefont {G.}~\bibnamefont
  {Tino}}\ and\ \bibinfo {editor} {\bibfnamefont {M.}~\bibnamefont
  {Kasevich}}}\ (\bibinfo  {publisher} {IOS Press, Amsterdam},\ \bibinfo {year}
  {2014})\ pp.\ \bibinfo {pages} {691--741},\ \Eprint
  {https://arxiv.org/abs/arXiv:1411.5164} {arXiv:1411.5164} \BibitemShut
  {NoStop}%
\bibitem [{\citenamefont {T\'oth}\ and\ \citenamefont
  {Apellaniz}(2014)}]{Toth2014Quantum}%
  \BibitemOpen
  \bibfield  {author} {\bibinfo {author} {\bibfnamefont {G.}~\bibnamefont
  {T\'oth}}\ and\ \bibinfo {author} {\bibfnamefont {I.}~\bibnamefont
  {Apellaniz}},\ }\bibfield  {title} {\bibinfo {title} {Quantum metrology from
  a quantum information science perspective},\ }\href
  {https://doi.org/10.1088/1751-8113/47/42/424006} {\bibfield  {journal}
  {\bibinfo  {journal} {J. Phys. A: Math. Theor.}\ }\textbf {\bibinfo {volume}
  {47}},\ \bibinfo {pages} {424006} (\bibinfo {year} {2014})}\BibitemShut
  {NoStop}%
\bibitem [{\citenamefont {Pezz\`e}\ \emph {et~al.}(2018)\citenamefont
  {Pezz\`e}, \citenamefont {Smerzi}, \citenamefont {Oberthaler}, \citenamefont
  {Schmied},\ and\ \citenamefont {Treutlein}}]{Pezze2018Quantum}%
  \BibitemOpen
  \bibfield  {author} {\bibinfo {author} {\bibfnamefont {L.}~\bibnamefont
  {Pezz\`e}}, \bibinfo {author} {\bibfnamefont {A.}~\bibnamefont {Smerzi}},
  \bibinfo {author} {\bibfnamefont {M.~K.}\ \bibnamefont {Oberthaler}},
  \bibinfo {author} {\bibfnamefont {R.}~\bibnamefont {Schmied}},\ and\ \bibinfo
  {author} {\bibfnamefont {P.}~\bibnamefont {Treutlein}},\ }\bibfield  {title}
  {\bibinfo {title} {Quantum metrology with nonclassical states of atomic
  ensembles},\ }\href {https://doi.org/10.1103/RevModPhys.90.035005} {\bibfield
   {journal} {\bibinfo  {journal} {Rev. Mod. Phys.}\ }\textbf {\bibinfo
  {volume} {90}},\ \bibinfo {pages} {035005} (\bibinfo {year}
  {2018})}\BibitemShut {NoStop}%
\bibitem [{\citenamefont {Beatty}\ and\ \citenamefont
  {Stilck~Fran{\c{c}}a}(2026)}]{Beatty2026Order}%
  \BibitemOpen
  \bibfield  {author} {\bibinfo {author} {\bibfnamefont {E.}~\bibnamefont
  {Beatty}}\ and\ \bibinfo {author} {\bibfnamefont {D.}~\bibnamefont
  {Stilck~Fran{\c{c}}a}},\ }\bibfield  {title} {\bibinfo {title} {Order $p$
  quantum {Wasserstein} distances from couplings},\ }\href
  {https://doi.org/10.1007/s00023-025-01557-z} {\bibfield  {journal} {\bibinfo
  {journal} {Ann. Henri Poincar\'e}\ }\textbf {\bibinfo {volume} {27}},\
  \bibinfo {pages} {787} (\bibinfo {year} {2026})}\BibitemShut {NoStop}%
\bibitem [{\citenamefont
  {Borsoni}(2026)}]{borsoni2026foldedoptimaltransportapplication}%
  \BibitemOpen
  \bibfield  {author} {\bibinfo {author} {\bibfnamefont {T.}~\bibnamefont
  {Borsoni}},\ }\bibfield  {title} {\bibinfo {title} {Folded optimal transport
  and its application to separable quantum optimal transport},\ }\href
  {https://arxiv.org/abs/2512.01722} {\bibfield  {journal} {\bibinfo  {journal}
  {arXiv:2512.01722}\ } (\bibinfo {year} {2026})}\BibitemShut {NoStop}%
\bibitem [{\citenamefont {T{\'o}th}\ and\ \citenamefont
  {Pitrik}(2025)}]{Toth2025QuantumWasserstein}%
  \BibitemOpen
  \bibfield  {author} {\bibinfo {author} {\bibfnamefont {G.}~\bibnamefont
  {T{\'o}th}}\ and\ \bibinfo {author} {\bibfnamefont {J.}~\bibnamefont
  {Pitrik}},\ }\bibfield  {title} {\bibinfo {title} {Quantum {Wasserstein}
  distance and its relation to several types of fidelities},\ }\href
  {https://arxiv.org/abs/2506.14523} {\bibfield  {journal} {\bibinfo  {journal}
  {arXiv:2506.14523}\ } (\bibinfo {year} {2025})}\BibitemShut {NoStop}%
\bibitem [{\citenamefont
  {Miller}(2026)}]{miller2026metricityseparablequantumoptimal}%
  \BibitemOpen
  \bibfield  {author} {\bibinfo {author} {\bibfnamefont {T.}~\bibnamefont
  {Miller}},\ }\bibfield  {title} {\bibinfo {title} {Metricity of separable
  quantum optimal transport with the {Hilbert-Schmidt} cost},\ }\href
  {https://arxiv.org/abs/2609.03769} {\bibfield  {journal} {\bibinfo  {journal}
  {arXiv:2609.03769}\ } (\bibinfo {year} {2026})}\BibitemShut {NoStop}%
\bibitem [{Note1()}]{Note1}%
  \BibitemOpen
  \bibinfo {note} {We use the convention that in the expression within the
  trace in Eq.~\protect \eqref {eq:GMPC_distance}, $H_n\otimes \protect
  \openone $ denotes an operator in which $H_n$ acts on subsystem $1$ and
  $\protect \openone $ acts on subsystem $2$}\BibitemShut {NoStop}%
\bibitem [{\citenamefont {T{\'o}th}\ and\ \citenamefont
  {Pitrik}(2026)}]{toth2026general}%
  \BibitemOpen
  \bibfield  {author} {\bibinfo {author} {\bibfnamefont {G.}~\bibnamefont
  {T{\'o}th}}\ and\ \bibinfo {author} {\bibfnamefont {J.}~\bibnamefont
  {Pitrik}},\ }\bibfield  {title} {\bibinfo {title} {General method for
  obtaining the energy minimum of spin {Hamiltonians} for separable states},\
  }\href {https://doi.org/10.48550/arXiv.2605.03022} {\bibfield  {journal}
  {\bibinfo  {journal} {arXiv:2605.03022}\ } (\bibinfo {year}
  {2026})}\BibitemShut {NoStop}%
\end{thebibliography}%

\end{document}